\documentclass[aps, prl, twocolumn, superscriptaddress, showpacs]{revtex4}
\usepackage{bbm}
\usepackage{mathrsfs}
\usepackage{amsfonts}
\usepackage{amsmath}
\usepackage{amssymb}
\usepackage{times}
\usepackage{graphicx}
\usepackage{bm}
\usepackage[colorlinks, linkcolor=blue, anchorcolor=blue, citecolor=red]{hyperref}

\newcommand\B{\rule[-1.0ex]{0pt}{0pt}}

\begin{document}
\title{Electron Dynamics in Slowly Varying Antiferromagnetic Texture}
\author{Ran Cheng}
\email{rancheng@physics.utexas.edu}
\affiliation{Department of Physics, University of Texas at Austin, Austin, Texas 78712, USA}
\author{Qian Niu}
\affiliation{Department of Physics, University of Texas at Austin, Austin, Texas 78712, USA}
\affiliation{International Center for Quantum Materials, Peking University, Beijing 100871, China}

\begin{abstract}
   Effective dynamics of conduction electrons in antiferromagnetic (AFM) materials with slowly varying spin texture is developed via non-Abelian gauge theory. Quite different from the ferromagnetic (FM) case, the spin of a conduction electron does not follow the background texture even in the adiabatic limit due to the accumulation of a SU(2) non-Abelian Berry phase. Correspondingly, it is found that the orbital dynamics becomes spin-dependent and is affected by two emergent gauge fields. While one of them is the non-Abelian generalization of what has been discovered in FM systems, the other leads to an anomalous velocity that has no FM counterpart. Two examples are provided to illustrate the distinctive spin dynamics of a conduction electron.
\end{abstract}

\maketitle

Interplay between current and magnetization is an essential issue underpinning the field of spintronics~\cite{ref:spintronics}, which is based primarily on the exchange interaction that couples the spins of conduction electrons and localized magnetic moments. In ferromagnetic (FM) materials with spin texture, the exchange interaction forces the conduction electrons to adjust their spins to local magnetization in the adiabatic limit where the background texture varies slowly over space-time. This induces a fictitious gauge field $\mathcal{F}_{\mu\nu}=\frac{\hbar}4\bm{n}\cdot(\partial_\mu\bm{n}\times\partial_\nu\bm{n})$ to the orbital dynamics of conduction electrons~\cite{ref:Volovik,ref:ZhangShoucheng,ref:Karin}, where $\bm{n}=\bm{M}/|\bm{M}|$ is the direction of local magnetization and $\partial_\mu$ denotes space-time derivatives. As a consequence, the interaction between current and the FM texture is recast into an emergent electrodynamics, and many celebrated phenomena should thereby be easily understood. For example, the spin motive force~\cite{ref:SMF} and the topological Hall effect~\cite{ref:THE} are well explained by the electric and magnetic components of the Lorentz force exerted by the gauge field; the back reaction of the Lorentz force in turn provides an intuitive interpretation to the current-induced spin torque exerted on the magnetic texture~\cite{ref:ZhangShoucheng,ref:Karin,ref:STT}.

However, the above picture apparently fails in antiferromagnetic (AFM) materials, where an itinerant spin finds no way to follow the orientation of the local moments that alters within unit cells. Nevertheless, we can define a staggered order parameter $\bm{n}=(\bm{M}_A-\bm{M}_B)/2M_s$ which can be slowly varying over space-time, where $\bm{M}_A$ and $\bm{M}_B$ are the alternating local moments and $M_s$ denotes their magnitudes. A natural question arises is how (or one step back, whether) this staggered order affects the dynamics of conduction electrons in the adiabatic limit. Despite recent theoretical~\cite{ref:AFMTheory} and experimental~\cite{ref:AFMExperiment} progress on AFM spintronics, this problem has never been addressed analytically. But for a full knowledge of spin transport in AFM materials, an analytic effective theory of the conduction electrons is desired. This motivates us to fill the need by generalizing the gauge theory that has been set up in FM systems.

In this Letter, we find that in a slowly varying AFM texture, instead of strictly following the local staggered order, the spin of a conduction electron obeys a linear differential equation~\eqref{eq:ds} which reflects an internal dynamics between degenerate bands in the adiabatic limit. Correspondingly, the orbital dynamics becomes spin-dependent and is influenced by two different emergent gauge fields. While one of them is shown to be the non-Abelian generalization of what would be responsible for the Lorentz force in the FM case, the other leads to an anomalous velocity which is truly new and unique to AFM systems. We have achieved this goal by applying the non-Abelian gauge theory~\cite{ref:Dimi,ref:NABerryPhase,ref:Dalibard} on a doubly degenerate band, where an underlying SU(2) non-Abelian Berry phase~\cite{ref:NABerryPhase,ref:BerryPhase} is responsible for the internal dynamics. With two examples demonstrating the novel properties of the spin dynamics at the end, this paper provides a general framework on how an AFM texture affects its conduction electrons. The converse problem, \emph{i.e.}, the back reaction of conduction electrons on the AFM background will be left for future inquiries.

Formalism -- Consider an AFM system on a bipartite lattice with the staggered order parameter $\bm{n}(\bm{r},t)$ varying slowly over space-time so that the system maintains local periodicity. $\bm{n}(\bm{r},t)$ will be treated separately from the conduction electrons and regarded as a given field. The spin of a conduction electron couples to the local moments by the exchange interaction $J(\bm{M}/M_s)\cdot\bm{\sigma}$, where $\bm{\sigma}$ is the vector of Pauli matrices standing for the spin operator (we set $\hbar=1$), and $\bm{M}$ flips sign on neighboring $A$ and $B$ sublattice sites. Accordingly, the conduction electron is described by a nearest neighbor tight-binding Hamiltonian locally defined around $\bm{n}(\bm{r},t)$ which is purely general,
\begin{align}
 \mathcal{H}(\bm{n}(\bm{r},t))=
 \begin{bmatrix}
 -J\bm{n}\!\cdot\!\bm{\sigma} \ \ & \gamma(\bm{k})\\
 \gamma^*(\bm{k}) & \ J\bm{n}\!\cdot\!\bm{\sigma}
 \end{bmatrix}
\end{align}
where $\gamma(\bm{k})=-t\sum_{\bm{\delta}}e^{i\bm{k}\cdot\bm{\delta}}$ is the hopping term with $\bm{\delta}$ connecting nearest neighboring $A-B$ sites. In general, $J$ can be negative if the exchange coupling is antiferromagnetic, but we assume a positive $J$ throughout this paper.

\begin{figure}[t]
   \centering
   \includegraphics[height=0.13\textheight]{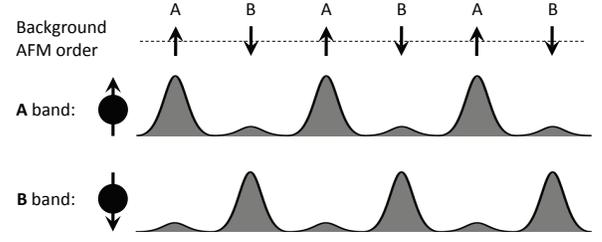}
   \caption{A schematic view of the Bloch waves in the lower band. Sub-band $A$ means a local spin up electron has a larger probability on the $A$ sites and a smaller probability on the $B$ sites; sub-band $B$ means the opposite case. They are degenerate in energy and their wave functions have a finite overlap depending on the ratio of $J/\varepsilon$.}{\label{Fig:SDW}}
\end{figure}

The local band structure can be easily solved as $\pm\varepsilon (\bm{k})$ with $\varepsilon(\bm{k})=\sqrt{J^2+|\gamma(\bm{k})|^2}$, and in the adiabatic limit we neglect transitions between $\varepsilon$ and $-\varepsilon$. Without loss of generality we will focus on the lower band $-\varepsilon$ which is doubly degenerate with the two sub-bands labeled by $A$ and $B$ (See Fig.~\ref{Fig:SDW}), whose wave functions are $|\psi_a\rangle=e^{i\bm{k}\cdot\bm{r}}|u_a\rangle$ and $|\psi_b\rangle=e^{i\bm{k}\cdot\bm{r}}|u_b\rangle$. The Bloch waves $|u_a\rangle=|A(\bm{k})\rangle|\!\uparrow(\bm{r},t)\rangle$ and $|u_b\rangle=|B(\bm{k})\rangle|\!\downarrow(\bm{r},t)\rangle$ maintain local periodicity around the space-time point $(\bm{r},t)$, where $|\!\uparrow(\bm{r},t)\rangle$ and $|\!\downarrow(\bm{r},t)\rangle$ are eigenstates of $\bm{n}\cdot\bm{\sigma}$. The periodic parts $|A(\bm{k})\rangle$ and $|B(\bm{k})\rangle$ exhibit opposite spatial patterns, which can be schematically understood in Fig.~\ref{Fig:SDW} that different spins try to find alternating sites so as to be aligned with the local moments. In the semiclassical point of view, an individual electron is described by a wave packet $|W\rangle=\int\mathrm{d}\bm{k}w(\bm{k})[\eta_a|\psi_a\rangle +\eta_b|\psi_b\rangle]$, where $\int\mathrm{d}\bm{k}\bm{k}w^2(\bm{k})=\bm{k}_c$ gives the center of mass momentum, and $\langle W|\bm{r}|W\rangle=\bm{r}_c$ is the center of mass position~\cite{ref:Dimi}. The coefficients $\eta_a$ and $\eta_b$ reflect the relative contributions from the two sub-bands and are constrained by $|\eta_a|^2+|\eta_b|^2=1$. While $\langle \psi_a|\psi_b\rangle=0$ due to the orthogonality of the spin eigenstates, $\langle A(\bm{k})|B(\bm{k})\rangle$ does not vanish. This finite overlap is of central importance to our theory so we hereby write it as,
\begin{align}
  \xi(\bm{k})\!=\!\langle A(\bm{\bm{k}})|B(\bm{k})\rangle\!=\!\frac{|\gamma(\bm{k})|}{\sqrt{J^2+|\gamma(\bm{k})|^2}}\!=\!\frac{\sqrt{\varepsilon^2-J^2}}{\varepsilon}. \label{eq:overlap}
\end{align}
It reaches maximum at the Brillouin zone (BZ) center and vanishes at the BZ boundary. It is obvious from Eq.~\eqref{eq:overlap} that the larger the $J$, the smaller the $\xi(\bm{k})$. If $J$ tends to infinity, the overlap $\xi(\bm{k})$ will vanish and the two sub-bands will be effectively decoupled. In this limit, the system becomes a simple combination of two independent FM sub-systems. In addition, $\xi(\bm{k})$ is a system parameter determined by the band structure alone, and it is conserved since from Eq.~\eqref{eq:overlap} we know that the energy conservation $\dot{\varepsilon}=0$ requires $\dot{\xi}=0$.

To construct effective theory on the degenerate band $-\varepsilon$, it is imperative to invoke the non-Abelian formalism~\cite{ref:Dimi,ref:NABerryPhase,ref:Dalibard} where dynamics between the $A$ and $B$ sub-bands introduces internal degree of freedom represented by the iso-spin vector $\bm{\eta}=\{\eta_1, \eta_2, \eta_3\}=\tilde{\eta}^\dagger\bm{\tau}\tilde{\eta}$, where $\tilde{\eta}=[\eta_a,\eta_b]^\mathrm{T}$. Here $\bm{\tau}$ is also a vector of Pauli matrices, but the different notation is adopted to avoid confusions with the spin operator $\bm{\sigma}$. The equations of motion of the electron wave packet are obtained through the effective Lagrangian approach~\cite{ref:Dimi} as,
\begin{subequations}
\label{eq:EOM}
 \begin{align}
  \dot{\bm{\eta}}&=2\bm{\eta}\times(\bm{\mathcal{A}}^r_\mu\dot{r}_\mu +\bm{\mathcal{A}}^k_\mu\dot{k}_\mu), \label{eq:EOMeta}\\
  \dot{k}_\mu& = \partial^r_\mu\varepsilon +\bm{\eta}\cdot[\bm{\Omega}^{rr}_{\mu\nu}\dot{r}_\nu+\bm{\Omega}^{rk}_{\mu\nu}\dot{k}_\nu], \quad \label{eq:EOMk}\\
  \dot{r}_\mu& = -\partial^k_\mu\varepsilon -\bm{\eta}\cdot[\bm{\Omega}^{kr}_{\mu\nu}\dot{r}_\nu+\bm{\Omega}^{kk}_{\mu\nu}\dot{k}_\nu], \label{eq:EOMr}
 \end{align}
\end{subequations}
where $r_\mu=(t,\bm{r}_c)$, but $k_\mu$ has no temporal component, its spatial components are just $\bm{k}_c$. In Eqs.~\eqref{eq:EOM} the $\cdot$ and $\times$ denote scalar and cross products in the iso-spin vector space. Here the non-Abelian Berry curvatures $\bm{\Omega}$ are obtained from the gauge potentials $\bm{\mathcal{A}}$ on the $A$ and $B$ sub-bands, for instance,
\begin{subequations}
\begin{align}
  &[\bm{\mathcal{A}}^r_\mu\cdot\bm{\tau}]_{ij}=i\langle u_i|\partial^r_\mu|u_j\rangle \label{eq:Berrypotential}\\
  &\bm{\Omega}^{rr}_{\mu\nu}=\partial^r_\mu \bm{\mathcal{A}}^r_\nu-\partial^r_\nu\bm{\mathcal{A}}^r_\mu +2\bm{\mathcal{A}}^r_\mu\times\bm{\mathcal{A}}^r_\nu, \label{eq:rrcurvature}
\end{align}
\end{subequations}
where $i,j$ run between $a,b$. The connection in BZ $\bm{\mathcal{A}}^k_\mu$ vanishes so $\bm{\Omega}^{kk}_{\mu\nu}=0$~\cite{ref:supplementary}, but $\bm{\Omega}^{kr}_{\mu\nu}=-\bm{\Omega}^{rk}_{\nu\mu}=\partial^k_\mu\bm{\mathcal{A}}^r_\nu$ is non-zero due to the finite overlap. Since the two sub-bands have opposite spins (see Fig.~\ref{Fig:SDW}), variation of $\bm{\eta}$ implies spin mistracking with the local order $\bm{n}$, and it worths emphasizing that this is due to the non-Abelian nature of the problem rather than any non-adiabatic process. However, the iso-spin vector $\bm{\eta}$ itself is not gauge invariant thus does not correspond to a physical observable directly. We need to relate $\bm{\eta}$ to the real spin of the electron defined by $\bm{s}=\langle W|\bm{\sigma}|W\rangle$ (in unit of $1/2$) which is fully gauge invariant. After some sophisticated calculations (see~\cite{ref:supplementary}), we obtain our central results:
\begin{subequations}
\label{eq:EOMtotal}
 \begin{align}
  \dot{\bm{s}} &= (1-\xi^2)(\bm{s}\cdot\bm{n})\dot{\bm{n}}, \label{eq:ds}\\
  \dot{\bm{k}} &= -\frac12\bm{n}\cdot(\nabla\bm{n}\times\dot{\bm{s}}), \label{eq:dk}\\
  \dot{\bm{r}} &= -\partial_{\bm{k}}\varepsilon -\frac12(\bm{s}\times\bm{n})\cdot\dot{\bm{n}}\ \partial_{\bm{k}}\ln\xi,\qquad \label{eq:dr}
 \end{align}
\end{subequations}
where $\dot{\bm{n}}=\partial_t\bm{n}+(\dot{\bm{r}}\cdot\nabla)\bm{n}$, and we have omitted the subscript $c$ of $\bm{r}_c$ and $\bm{k}_c$ for convenience. We assert that Eqs.~\eqref{eq:EOMtotal} are the fundamental equations of motion of a conduction electron in an AFM texture, which are represented by the joint evolutions of three parameters $(\bm{s}, \bm{r}, \bm{k})$. An essential character that distinguishes the AFM electron dynamics from its FM counterpart lies in Eq.~\eqref{eq:ds}, from which we know that the real spin $\bm{s}$ of a conduction electron does not follow the order parameter $\bm{n}$ in the adiabatic limit. Eq.~\eqref{eq:ds} is purely geometrical because $\mathrm{d}t$ can be eliminated on both sides, therefore, a path of $\bm{n}$ is mapped onto a path of $\bm{s}$. The mistrack between $\bm{s}$ and $\bm{n}$ is due to inter-sub-band dynamics through the accumulation of a SU(2) non-Abelian Berry phase $\mathcal{P}\exp[-i\int\bm{\mathcal{A}}^r_\mu\cdot\bm{\tau}\mathrm{d}r_\mu]$ along the trajectory in real space~\cite{ref:NABerryPhase}. Remarkably, we note that the connection $\bm{\mathcal{A}}^r_\mu$ is the same as the model proposed by Ref.~\cite{ref:BPSmonopole} if $\xi$ is identified with their parameter $f(B)$. Subsequently, the SU(2) Berry phase can be attributed to the gauge flux of a 't Hooft-Polyakov monopole located at the origin of the parameter space spanned by $\bm{n}$.

\begin{figure}[t]
   \centering
   \includegraphics[height=0.13\textheight]{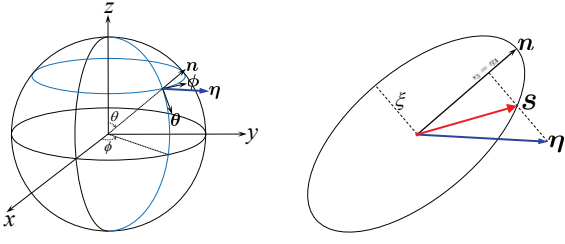}
   \caption{(Color online) Left panel: the iso-spin vector $\bm{\eta}$ (in blue) in the local frame that moves with $\bm{n}$: $\bm{\eta}=\eta_1\bm{\theta}+\eta_2\bm{\phi}+\eta_3\bm{n}$, where $\bm{\theta}$ and $\bm{\phi}$ are spherical unit vectors. Right panel: In our particular gauge, $\bm{\eta}$ is coplanar with $\bm{n}$ and $\bm{s}$ (in red). But no matter what gauge we choose, $s_3=\eta_3$ and $s_1^2+s_2^2=\xi^2(\eta_1^2+\eta_2^2)$ always hold. \label{Fig:spinfootball}}
\end{figure}

Discussions -- An important fact about non-Abelian theory is that the gauge fields $\bm{\Omega}$'s are not gauge invariant but the iso-spin scalars $\bm{\eta}\cdot\bm{\Omega}$ appearing in Eqs.~\eqref{eq:EOM} are. The gauge freedom originates from the phase ambiguity of the local spin wave functions which are obtained by acting $U(\bm{r}, t)=e^{-i\sigma_z\phi/2}e^{-i\sigma_y\theta/2}e^{-i\sigma_z\chi/2}$ on the eigenstates of $\sigma_z$. Here $\theta(\bm{r},t)$ and $\phi(\bm{r},t)$ are the spherical angles specifying the direction of $\bm{n}$, whereas $\chi(\bm{r}, t)$ can be chosen arbitrarily and is not physical, but in deriving Eqs.~\eqref{eq:EOMtotal} we have fixed the gauge by setting $\chi=0$. The gauge fields (Berry curvatures) not only drive the evolutions of $\bm{k}$ and $\bm{r}$, but also determine inter-sub-band evolutions described by the dynamics of $\bm{\eta}$, which has been converted to the dynamics of $\bm{s}$. To see the intrinsic relationship between $\bm{s}$ and $\bm{\eta}$ more explicitly, we now derive from Eq.~\eqref{eq:ds} the following expression (See~\cite{ref:supplementary}),
\begin{align}
  (\bm{s}\cdot\bm{n})^2+\frac{(\bm{s}\times\bm{n})^2}{\xi^2}=s_3^2+\frac{s_1^2+s_2^2}{\xi^2}=1, \label{eq:ellipsoid}
\end{align}
which indicates that the tip of $\bm{s}$ moves on an prolate spheroid with the semi-major axis being the local order parameter $\bm{n}$, and the semi-minor axis has length $\xi$. On the other hand, $\bm{\eta}$ is constrained by $\bm{\eta}^2=|\eta_a|^2+|\eta_b|^2=1$. In our particular gauge marked by $\chi=0$, $\bm{\eta}$ can be pictured as a vector in the local frame that moves with $\bm{n}(\bm{r}, t)$ hence $\bm{\eta}=\eta_1\bm{\theta}+\eta_2\bm{\phi}+\eta_3\bm{n}$ (Fig.~\ref{Fig:spinfootball}, left), meanwhile it is coplanar with $\bm{n}$ and $\bm{s}$ (Fig.~\ref{Fig:spinfootball}, right). For arbitrary gauges, some manipulations show that we always have $s_3=\eta_3$ and $s_1^2+s_2^2=\xi^2(\eta_1^2+\eta_2^2)$. A gauge transformation can only change the angles between $s_{1,2}$ and $\eta_{1,2}$, while $\eta_3$ is gauge invariant.

The above properties can be further understood from two aspects: \emph{(i)} Since opposite local spins are associated with different sublattice sites, we are able to infer the spin projection $s_3$ of an electron along $\bm{n}$ by measuring the probability difference $\eta_3=|\eta_a|^2-|\eta_b|^2$ on neighboring $A-B$ sites (vice versa). This explains why $\eta_3=s_3$ and thus it is a physical variable. In contrast, $\eta_1$ and $\eta_2$ are not physical due to the gauge freedom. \emph{(ii)} The reduced density matrix for the spin degree of freedom after tracing out the sublattice can be written as $\rho_s=\frac12(1+\bm{a}\cdot\bm{\sigma})$, thus $\bm{s}=\mathrm{Tr}[\rho_s\bm{\sigma}]=\bm{a}$. On the spheroid we have $s^2\le1$, and it results in $\mathrm{Tr}\rho_s^2\le\mathrm{Tr}\rho_s=1$, which suggests that the conduction electron is effectively in a mixed spin state. This again can be attributed to the locking of $s_3$ and $\eta_3$, which means the spin and sublattice degrees of freedom are entangled. The entanglement provides information on the spin orientation from the knowledge of sublattice so it destroys the coherence of the spin states.

\begin{table}[t]
 \begin{tabular}{l|l}
  \hline\hline
  FM spin texture & AFM spin texture \B\\
  \hline
  $\bm{s}=\bm{n}$                                    & $\ \dot{\bm{s}}=(1-\xi^2)(\bm{s}\cdot\bm{n})\dot{\bm{n}}$\\
  U(1) Abelian Berry Phase: & SU(2) non-Abelian Berry Phase:\\
  $\gamma(\Gamma)=\oint_{\Gamma}\mathcal{A}_\mu\mathrm{d}r_\mu$ & $U(\Gamma)=\mathcal{P}\exp[-i\oint_{\Gamma}\bm{\mathcal{A}}^r_\mu\cdot\bm{\tau}\mathrm{d}r_\mu]$\\
  Dirac monopole & 't Hooft-Polyakov monopole\\
  \hline
  $\dot{\bm{k}}=\bm{E}+\bm{B}\times\dot{\bm{r}} \qquad\qquad\ \ \ $     & $\ \dot{\bm{k}}=(1-\xi^2)(\bm{s}\cdot\bm{n})(\bm{E}+\bm{B}\times\dot{\bm{r}})$\\
  $\dot{\bm{r}}=-\partial_{\bm{k}}\varepsilon$ \B     & $\ \dot{\bm{r}}=-\partial_{\bm{k}}\varepsilon -\frac12(\bm{s}\times\bm{n})\cdot\dot{\bm{n}}\ \partial_{\bm{k}}\ln\xi$\\
  \hline\hline
 \end{tabular}
 \caption{Comparison of the electron dynamics in FM and AFM spin textures. In the FM case, spin dynamics is trivial, and along a closed path $\Gamma$ the electron acquires an U(1) Berry phase which can be regarded as the magnetic flux of a Dirac monopole. A Lorentz force is resulted in the orbital motion. In the AFM case, spin dynamics is non-trivial due to the mixture of the two degenerate sub-bands through a SU(2) non-Abelian Berry phase, which originates from the gauge field generated by a 't Hooft-Polyakov monopole. Consequently, a spin-dependent Lorentz force and an anomalous velocity appear in the orbital dynamics. }{\label{Tab:comparison}}
\end{table}

Furthermore, we turn to the orbital dynamics which is spin-coupled. By substituting Eq.~\eqref{eq:ds} into~\eqref{eq:dk} we get,
\begin{align}
  \dot{\bm{k}}&=(1-\xi^2)(\bm{s}\cdot\bm{n})(\bm{E}+\bm{B}\times\dot{\bm{r}}), \label{eq:SMF} \\
        \bm{E}&=\frac{\sin\theta}2(\partial_t\theta\nabla\phi-\nabla\theta\partial_t\phi);\ \bm{B}=\frac{\sin\theta}2(\nabla\theta\times\nabla\phi), \label{eq:BE}
\end{align}
the $\bm{E}$ and $\bm{B}$ fields here are the same as what have been discovered in FM textures, where they are responsible for the spin motive force~\cite{ref:SMF} and the topological Hall effect~\cite{ref:THE}, respectively. However, quite different from the FM case, the gauge charge $\bm{s}\cdot\bm{n}$ in Eq.~\eqref{eq:SMF} is spin dependent, and the factor $\xi^2$ results from the non-commutative term $2\bm{\mathcal{A}}^r_\mu\times\bm{\mathcal{A}}^r_\nu$ in Eq.~\eqref{eq:rrcurvature}, they both reflect the influence of the spin dynamics on the orbital motion. The parameter $\xi\in(0,1)$ plays a key role here: in the $\xi\rightarrow1$ limit, $1-\xi^2$ vanishes thus from Eq.~\eqref{eq:ds} and Eq.~\eqref{eq:SMF} we get null results $\dot{\bm{s}}=0$ and $\dot{\bm{k}}=0$. In the other limit where $\xi\rightarrow0$, the solution of Eq.~\eqref{eq:ds} reduces to $\bm{s}=\pm\bm{n}$ upon the initial condition $\bm{s}(0)=\pm\bm{n}(0)$, thus Eq.~\eqref{eq:SMF} reduces to the usual Lorentz force equation, by which the system losses the manifest non-Abelian feature and behaves as two decoupled FM sub-systems. It deserves attention that although the spin dynamics becomes trivial in the $\xi\rightarrow0$ limit, a subtle difference still exists in the AFM system: the two sub-bands have opposite gauge charges since their spins are of opposite directions. Therefore, the Lorentz forces exerting on the electrons from the two sub-bands are of opposite signs, which may lead to spin-dependent transport.

Besides, the real space dynamics governed by Eq.~\eqref{eq:dr} also exhibits spin-orbit coupling through $\frac12(\bm{s}\times\bm{n})\cdot\dot{\bm{n}}\ \partial_{\bm{k}}\ln\xi$ which is an anomalous velocity. This term is along the same direction as $\partial_{\bm{k}}\varepsilon$, so Eq.~\eqref{eq:dr} amounts to give a modified band velocity. It worth mentioning that this term is unique to AFM systems and has nothing to do with the anomalous velocity studied in FM systems and quantum Hall systems~\cite{ref:Dimi}. It comes from the $\bm{\Omega}^{kr}_{\mu\nu}$ curvature that joints the real space with BZ, whose importance has been overlooked before. For better comparison, we summarize the fundamental electron dynamics of FM and AFM textures in Tab.~\ref{Tab:comparison}.

\begin{figure}[t]
   \centering
   \includegraphics[height=0.23\textheight]{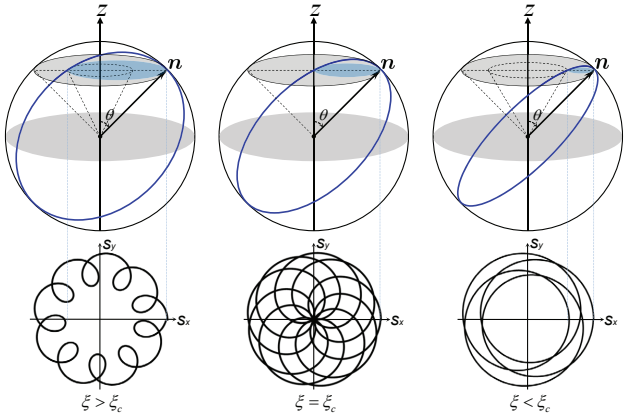}
   \caption{(Color online) Spin evolutions for three different $\xi$'s when $\bm{n}(t)$ is moving round a cone with constant angle $\theta$ from the $z$-axis. Upper panels: the tip of $\bm{s}$ is confined on the intersection of the cone's bottom and the spheroid described by Eq.~\eqref{eq:ellipsoid}. Lower panels: orbits of the tip plotted with $s_x$ and $s_y$. The topology of the orbits is separated into two classes (left and right) by the critical case (middle) where $\xi_c^2=\cos^2\theta/(1+\cos^2\theta)$ and the inner cone shrinks to zero. The orbits are not necessarily commensurate with $\bm{n}$.} {\label{Fig:Rings}}
\end{figure}

Examples -- First consider an electron passing through an AFM domain wall of either the Bloch type or the N\'{e}el type. We assume that on the incident side $\bm{s}$ is polarized along the staggered order parameter $\bm{n}$. If it were a FM domain wall then the total spin rotation on the outgoing side is a topological quantity $\Delta \varphi_s=\pm\pi$ where $+1(-1)$ is the topological number of the (anti-) domain wall. Here for an AFM domain wall, by solving Eq.~\eqref{eq:ds} it is found that the spin rotation is quantized by a renormalized unit: $\Delta \varphi_s=\pm\Pi$ where
\begin{align}
  \Pi=\left\{
  \begin{array}{cl}
   \pi-\arctan[\xi\tan\xi\pi] & \mbox{if $\xi\!<\!\frac12$} \\
    -\arctan[\xi\tan\xi\pi] & \mbox{if $\xi\!>\!\frac12$},
  \end{array}
  \right.
\end{align}
in the $\xi\rightarrow0$ limit $\Pi$ reduces to $\pi$, and in the $\xi\rightarrow\infty$ limit $\Pi$ vanishes. Note $\Pi$ only depends on the system parameter $\xi$.

Consider a second case where $\bm{n}(t)$ is varying round a cone of constant semi-angel $\theta$ in the lab frame (see Fig.~\ref{Fig:Rings}), which can be realized in a spin wave. According to Eq.~\eqref{eq:ds} we know that $\mathrm{d}s_z=0$ due to $\mathrm{d}n_z=0$, thus the tip of $\bm{s}$ should stay in the bottom plane of the cone. On the other hand, we learn from Eq.~\eqref{eq:ellipsoid} that the tip is constrained on the spheroid that moves with the instantaneous $\bm{n}(t)$. Therefore, the actual orbit traversed by the tip is contained in the intersection of the two constraints, and the spin $\bm{s}$ is bounded between the $\bm{n}$-cone and an inner cone whose semiangle depends on the system parameter $\xi$. Depicted in Fig.~\ref{Fig:Rings}, we find that the motion of $\bm{s}$ in the lab frame exhibits both precession and nutation, which is reflected by the orbits of the tip. By tuning the parameter $\xi$, the motion of $\bm{s}$ falls into two topologically distinct classes separated by the critical condition $\xi_c^2=\cos^2\theta/(1+\cos^2\theta)$ or simply $|\gamma(\bm{k})|=J\cos\theta$.

\emph{Acknowledgements} - The authors are grateful to Yizhuang You and Biao Wu for constant helps. We thank K.~Everschor, J.~Zhou, X.~Li, D.~Xiao, S.~A.~Yang, Y.~Gao, and A.~H.~MacDonald for insightful comments. This work is supported by NSF, DOE, the Welch foundation, and NFSC.


\begin{thebibliography}{32}
  \bibitem{ref:spintronics} I. \v{Z}uti\'{c}, J. Fabian, and S. D. Sarma, Rev. Mod. Phys. \textbf{76}, 323 (2004) and the reference therein.
  \bibitem{ref:Volovik} G. E. Volovik, J. Phys. C \textbf{20}, L83 (1987).
  \bibitem{ref:ZhangShoucheng} Y. B. Bazaliy, B. A. Jones, and S. -C. Zhang, Phys. Rev. B \textbf{57}, R3213 (1998).
  \bibitem{ref:Karin} T. Schulz \emph{et al.}, Nature Phys. doi:10.1038/nphys2231 (2012); K. Everschor, M. Garst, R. A. Duine, and A. Rosch, Phys. Rev. B \textbf{84}, 064401 (2011).
  \bibitem{ref:SMF} S. A. Yang \emph{et al.}, Phys. Rev. Lett. \textbf{102}, 067201 (2009); S. E. Barnes and S. Maekawa, Phys. Rev. Lett. \textbf{98}, 246601 (2007);
  \bibitem{ref:THE} M. Lee \emph{et al.}, Phys. Rev. Lett. \textbf{102}, 186601(2009); A. Neubauer \emph{et al.}, Phys. Rev. Lett. \textbf{102}, 186602 (2009); P. Bruno, V. K. Dugaev, and M. Taillefumier, Phys. Rev. Lett. \textbf{93}, 096806 (2004); J. Ye \emph{et al.}, Phys. Rev. Lett. \textbf{83}, 3737 (1999).
  \bibitem{ref:STT} J. Zang, M. Mostovoy, J. H. Han, and N. Nagaosa, Phys. Rev. Lett. \textbf{107}, 136804 (2011); F. Jonietz \emph{et al.}, Science \textbf{330}, 1648 (2010).
  \bibitem{ref:AFMTheory} A. C. Swaving and R. A. Duine, J. Phys.: Cond. Mat. \textbf{24}, 024223 (2012); K. M. D. Hals, Y. Tserkovnyak, and A. Brataas, Phys. Rev. Lett. \textbf{106}, 107206 (2011); R. Wieser, E. Y. Vedmedenko, and R.~Wiesendanger, Phys. Rev. Lett. \textbf{106}, 067204 (2011); A. C. Swaving and R. A. Duine, Phys. Rev. B \textbf{83}, 054428 (2011); Y. Xu, S. Wang, and K. Xia, Phys. Rev. Lett. \textbf{100}, 226602 (2008); P. M. Haney and A. H. MacDonald, Phys. Rev. Lett. \textbf{100}, 196801 (2008).
  \bibitem{ref:AFMExperiment} S. Urazhdin and N. Anthony, Phys. Rev. Lett. \textbf{99}, 046602 (2007); R. Jaramillo \emph{et al.}, Phys. Rev. Lett. \textbf{98}, 117206 (2007); Z. Wei \emph{et al.}, Phys. Rev. Lett. \textbf{98}, 116603 (2007); A. H. MacDonald and M. Tsoi, Phil. Trans. R. Soc. A \textbf{369}, 3098 (2011).
  \bibitem{ref:Dimi} D. Culcer, Y. Yao, and Q. Niu, Phys. Rev. B \textbf{72}, 085110 (2005); D. Xiao, M. -C. Zhang, and Q. Niu, Rev. Mod. Phys. \textbf{82}, 1959 (2010) and the reference therein.
  \bibitem{ref:Dalibard} J. Dalibard, \emph{et al.}, Rev. Mod. Phys. \textbf{83}, 1523 (2011).
  \bibitem{ref:NABerryPhase} F. Wilczek and A. Zee, Phys. Rev. Lett. \textbf{52}, 2111 (1984); J. Moody, A. Shapere, and F. Wilczek, Phys. Rev. Lett. \textbf{56}, 893 (1986); C. A. Mead, Phys. Rev. Lett. \textbf{59}, 161 (1987); B. Zygelman, Phys. Rev. Lett. \textbf{64}, 256 (1990).
  \bibitem{ref:BerryPhase} M. V. Berry, Proc. R. Soc. London. A \textbf{392}, 45 (1984).
  \bibitem{ref:supplementary} See the supplementrary materials of this paper.
  \bibitem{ref:BPSmonopole} J. Sonner and D. Tong, Phys, Rev, Lett. \textbf{102}, 191801 (2009).
\end{thebibliography}
\end{document}